\documentclass[aps,pra,twocolumn,showpacs,superscriptaddress]{revtex4}
\usepackage{graphicx}
\usepackage{dcolumn}
\usepackage{color}
\usepackage{bm}
\usepackage{amssymb}
\usepackage{amsmath}
\begin{document}
\title{A toroidal trap for the cold $^{87}Rb$ atoms using a rf-dressed quadrupole trap}
\author{A. Chakraborty}
\email[E-mail: ]{carijit@rrcat.gov.in}
\author{S. R. Mishra}
\affiliation{Homi Bhabha National Institute, Mumbai-400094, India.}
\affiliation{Raja Ramanna Centre for Advanced Technology, Indore-452013, India.}
\author{S. P. Ram}
\author{S. K. Tiwari}
\author{H. S. Rawat}
\affiliation{Raja Ramanna Centre for Advanced Technology, Indore-452013, India.}

\begin{abstract}
We demonstrate the trapping of cold $^{87}Rb$ atoms in a toroidal geometry using a rf-dressed quadrupole magnetic trap formed by superposing a strong radio frequency (rf) field on a quadrupole trap. This rf-dressed quadrupole trap has minimum of the potential away from the quadrupole trap centre on a circular path which facilitates the trapping in the toroidal geometry. In the experiments, the laser cooled atoms were first trapped in the quadrupole trap, then cooled evaporatively using a weak rf-field, and finally trapped in the rf-dressed quadrupole trap. The radius of the toroid could be varied by varying the frequency of the dressing rf-field. It has also been demonstrated that a single rf source and an antenna can be used for the rf-evaporative cooling as well as for rf-dressing of atoms. The atoms trapped in the toroidal trap may have applications in realization of an atom gyroscope as well as in studying the quantum gases in low dimensions.
\end{abstract}
\pacs{03.75.Be, 37.10.Gh, 05.30.Jp, 67.85.-d, 39.25.+k}
\maketitle

\section{Introduction}

The development of atom traps with sophisticated potential landscapes has catalysed research in the field of cold quantum gases. Optical lattices \cite{Windpassinger:2013}, double-well \cite{Hofferberth:2007} and potentials confining atoms in low dimensions (two dimensions (2D) and one dimension (1D)) \cite{Merloti:2013} are examples where new physics has been elucidated in the recent past. Trapping atoms in low dimensions is of interest for several reasons \cite{Petrov:2004}. The phase transition to Bose-Einstein condensation (BEC), which is forbidden for a homogeneous Bose gas in 1D and 2D geometries, becomes feasible if confining 1D or 2D potentials obey the suitable power laws for spatial variation. The atom trapping in toroidal or ring shaped geometry is useful for the study of coherence, super-fluidity, Josephson oscillations etc of the confined quantum gases \cite{Bloch:1973,Albiez:2005} in low dimensions, besides having the applications in the realization of an atom gyroscope.

The proposals and realizations of the ring shaped atom traps have been reported by several groups in the recent past employing various techniques. The use of static magnetic traps superimposed with a strong rf-field, called rf-dressed magnetic traps, appear promising due to flexibility and control available on the generated trapping potentials \cite{Zobay:2001}. The reports on ring shaped traps include the use of a specially designed magnetic coils for ring shaped quadrupole trap \cite{Fernholz:2007}, a rf-dressed magnetic trap in combination with a standing wave pattern of an optical beam \cite{Morizot:2006}, and a rf-dressed magnetic trap in combination with a sheet type dipole laser beam \cite{Heathcote:2008}. In an interesting work, using the hollow beams and an optical lattice, \citet{Amico:2009} have proposed and demonstrated the generation of ring optical lattice for trapping of the superfluid atomic systems to use as qubits.

In the rf-dressed magnetic trap, an atom experiences the adiabatic potential which is position dependent eigen-energy of the dressed-state of the atom while it interacts with the static and oscillating magnetic fields. This type of adiabatic potentials, also referred as rf-dressed potentials, can be handled with less complexity as compared to the potentials of optical beams. These adiabatic potentials also permit ample control over the generated potential landscape which can be tuned by tuning the amplitude, frequency and phase of the time varying rf fields \cite{Folman:2002,Colombe:2004,Lesanovsky:2006:73,Morizot:2006,Morizot:2007,Sherlock:2011,Lesanovsky:2007,Merloti:R:2013,Morizot:2008,Gildemeister:2010}. Depending upon the rf field parameters, the rf-dressed potentials can offer exquisite trapping geometries like multiple-wells, asymmetric arcs, ring traps and rotating (or oscillating) toroidal traps \cite{Chakraborty:2014}. Such non-trivial trapping geometries are considered important in conventional as well as miniaturised atom traps \cite{Schumm:2005} for various applications.

In this work, we demonstrate the trapping of cold $^{87}Rb$ atoms in a toroidal geometry using a rf-dressed quadrupole magnetic trap. As compared to the earlier approaches, in which the toroidal trap was formed either using a rf-dressed magnetic trap and dipole potential of a laser beam \cite{Heathcote:2008} or using a specially designed magnetic coils for ring shaped quadrupole traps \cite{Fernholz:2007}, our method is simple to implement as it requires only a rf-dressed quadrupole magnetic trap. In our method, the atom-cloud trapped in the quadrupole trap in $|F=2, m_F=2\rangle$ state is first exposed to a weak rf field (with frequency sweep) for evaporative cooling, and then the stronger rf-field at a different frequency is used to transfer the cloud to the rf-dressed potential. The rf-field used for dressing is also be subjected to a small range frequency sweep (few MHz) to obtain a clean toroidal shape of the trapped cloud. The observed life-time of atoms in rf-dressed potential is longer than that in a bare quadrupole trap due to suppression of Majorana flips in the rf-dressed potential. It is also demonstrated that a single rf source and antenna can be used for the evaporative cooling as well as for the rf-dressing to trap atoms in the toroidal geometry. 

In this work, it is further experimentally demonstrated that use of an additional rf-field along the quadrupole trap axis, in presence of the linearly polarized rf-field perpendicular to the trap axis, gives rises to asymmetry in the trapped atom cloud on the ring, with the cloud being held at a particular position on the ring. These results are in agreement with the theoretical predictions. Since the position of trapped cloud can be altered by altering the phase of the axial rf-field \cite{Chakraborty:2014}, the modulation of the phase of the field can provide an opportunity to rotate the cloud along the ring. As compared to use of a dipole laser to stir the cloud, this may be a simple and robust method to rotate the atom cloud on the ring and may be useful to study super-fluidity of ultra-cold atomic sample in the ring geometry. These experiments will be attempted by us in the future.

The article is organized as follows. In the section \ref{Toroidal Radio Frequency Dressed Potentials}, the theoretical background required for the rf-dressed potentials is discussed. The section \ref{Experimental Realisation} gives the description of the experimental setup used for the generation of the toroidal rf-dressed potentials in the present work. The main results of the work are discussed in the section \ref{Results}. Finally, the conclusions of the present work are given in the section \ref{Conclusion}.

\section{RF-dressed quadrupole trap}\label{Toroidal Radio Frequency Dressed Potentials}

\begin{figure}
\includegraphics[width=8.6cm]{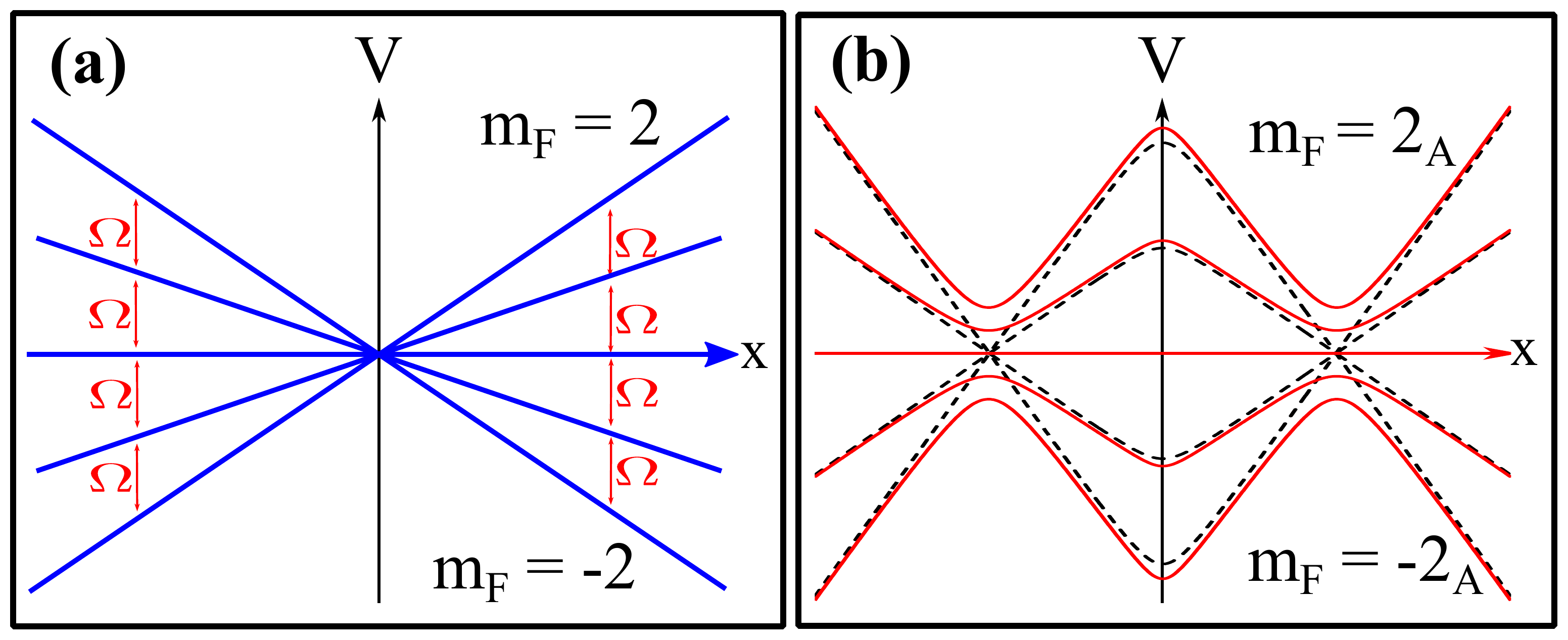}
\caption{\label{fig:field_line}(Color online) Variation in potential energy with position for an atom: (a) in bare quadrupole trap and (b) in rf-dressed quadrupole trap. The potential in graph (b) (continuous curve) shows the avoided crossing at the regions of resonance due to rf-field induced coupling between the states. }
\end{figure}

Trapping of neutral atoms in an inhomogeneous static magnetic field superimposed with a time varying rf-field can be well described using a semi-classical approach \cite{Heathcote:thesis}. The rf-field couples various hyperfine Zeeman sub-levels of an atom and the rf-dressed potential is the effective potential experienced by the atom in static and time varying fields. The rf-dressed potential is also an energy eigen value of the Hamiltonian in the dressed state picture. Figure \ref{fig:field_line} shows a schematic diagram of the effect of a rf-field on the potential energy of an atom trapped in a quadrupole trap. After rf dressing, the potential minimum shifts away from the quadrupole trap centre and forms avoided crossing as shown in Fig. \ref{fig:field_line}(b). In order to evaluate the potential energy $V(r)$ of an atom in the static and rf-fields, we consider a quadrupole magnetic trap having the field distribution as,
\begin{equation}\label{eq:quadfield}
\textbf{B}^\textbf{S}\textbf{(r)}=B_q\left(\begin{array}{c}
x\\
y\\
-2z\\
\end{array}\right),
\end{equation}
where $B_q$ is the radial field gradient. We also consider the rf-field of the form $\textbf{B}^\textbf{{rf}}\textbf{(t)}=\{B_x\cos\omega t,B_y\cos(\omega t-\alpha),B_z\cos(\omega t-\beta)\}$ is used for the dressing purpose, where $B_x$, $B_y$ and $B_z$ are the amplitudes of the rf-field in three orthogonal directions, and, $\alpha$ and $\beta$ are the relative phases.

As is known that trapping in the bare quadrupole trap results in Zeeman splitting of hyperfine levels of $^{87}Rb$ atoms with transition frequency between adjacent levels given as, 
\begin{equation}
\omega_0=\frac{g_F\mu_BB_q}{\hbar}\sqrt{x^2+y^2+4z^2},
\end{equation}
where $g_F$ is the Lande's g-factor, $\mu_B$ is the Bohr magneton and $\hbar$ is the reduced Planck's constant. This transition frequency, known as Larmor frequency, is inherently position dependent and radially isotropic. With the rf radiation of frequency $\omega$, the transitions between adjacent Zeeman levels of a hyperfine level can be excited if $\omega\approx\omega_0$. Because of the position dependent nature of the transition frequency (i.e. Larmor frequency), a single frequency rf-source does not excite transitions at every position in the trap. This spatially varying transition frequency in the quadrupole trap contributes to the position dependent nature of the rf-dressed potential energy $V(r)$. For an atom in a Zeeman hyperfine sub-level $m_F$, the potential energy $V(r)$ can be calculated using a rotating wave approximation formalism \cite{Lesanovsky:2007,Heathcote:2008,Chakraborty:2014} and can be written including the gravity as
\begin{equation}\label{eq:pot_compact}
V=m_F\hbar\sqrt{\delta^2+|\Omega|^2}+mgy.
\end{equation}
where
\begin{equation}\label{eq:delta}
\delta=\omega-\omega_0,
\end{equation}
where $(\delta)$ is detuning of the dressing rf-radiation frequency from the Larmor frequency, $(\Omega)$ is Rabi frequency for coupling between the sub-levels $m_F$ and $m_F\pm1$. The Rabi frequency can be determined as
\begin{multline}\label{eq:omegafinal}
|\Omega|^2=\left(\frac{g_F\mu_B}{2\hbar}\right)^2 \Bigg[\frac{4z^2}{x^2+y^2+4z^2}\left( \frac{B_x^2x^2+B_y^2y^2}{x^2+y^2}\right)\\ +\left(\frac{B_x^2y^2+B_y^2x^2}{x^2+y^2}\right)+B_z^2\left( \frac{x^2+y^2}{x^2+y^2+4z^2}\right)\\-\frac{2B_xB_yxy\cos\alpha}{x^2+y^2+4z^2}+\frac{4B_xB_yz\sin\alpha}{\sqrt{x^2+y^2+4z^2}}\\+\frac{4B_yB_zyz\cos(\alpha-\beta)}{x^2+y^2+4z^2}+\frac{2B_yB_zx\sin(\alpha-\beta)}{\sqrt{x^2+y^2+4z^2}}\\+\frac{4B_zB_xzx\cos\beta}{x^2+y^2+4z^2}+\frac{2B_zB_xy\sin\beta}{\sqrt{x^2+y^2+4z^2}}\Bigg].
\end{multline}

\begin{figure}[t]
\includegraphics[width=8.6cm]{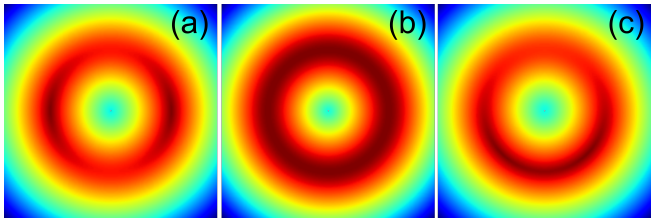}
\caption{\label{fig:simulation}(Color online) The calculated contours of rf-dressed potential V (Eq. (\ref{eq:pot_compact}), (\ref{eq:delta}) and (\ref{eq:omegafinal})) in the xy-plane for different values of parameters. The plot (a) shows a double-well potential with $B_x=0.7$ G, $B_y=B_z=0$, plot (b) shows the ring trap with $B_x=B_y=0.7$ G, $B_z=0$ and $\alpha=-\pi/2$, and plot (c) shows an asymmetric ring trap with $B_x=0.7$ G, $B_y=0$, $B_z=0.2$ G and $\beta=0$. The other parameters, $B_q$ = 100 G $cm^{-1}$, $\omega$= 2$\pi\times$1.5 MHz etc., are common to all the plots. The colours red, yellow, green and blue show the potential values in increasing order.}
\end{figure}

With the appropriate choice of the field strengths $B_q$, $B_x$, $B_y$, $B_z$ and phases $\alpha$ and $\beta$, variety of potentials can be generated \cite{Chakraborty:2014} and some of these are shown in Fig. \ref{fig:simulation}. The numerical simulations to determine the potential (from Eq. (\ref{eq:pot_compact})) and predict the trapping geometry take into account the contribution of the detuning $(\delta)$ and that of the Rabi frequency $(\Omega)$ associated with the rf-field. The simulations help in choosing the various parameters for the desired trap geometry by ensuring the potential depth suitable to trap an atom cloud at known temperature.

From Eq. (\ref{eq:pot_compact}), (\ref{eq:delta}) and (\ref{eq:omegafinal}), it can be determined that with a linear polarisation of rf-field (\textit{i.e.} $B_x\neq 0$, $B_y=B_z=0$), the rf-dressed potential is expected to be a double-well potential as shown in Fig. \ref{fig:simulation}(a). These wells are formed on the circumference of a ring such that their positions are symmetric along the ring circumference, when gravity is negligible. With a circular polarisation of rf-field (\textit{i.e.} $B_x\neq 0$, $B_y\neq 0$, $B_z=0$ with $\alpha=\pm\pi/2$), the resulting potential is expected to be a ring trap potential as shown in Fig. \ref{fig:simulation}(b). An addition of the rf-field polarized along the qradrupole trap axis (\textit{i.e.} $B_z\neq 0$), in presence of the linearly polarized rf-field (\textit{i.e.} $B_x\neq 0$), results in an asymmetric ring potential with its local minimum on a ring as shown in Fig. \ref{fig:simulation}(c). The spatial position of this local minimum is governed by the phase angle $\beta$ of the axial rf-field. 

The shape of the rf-dressed potentials may be affected by the gravity. The parameter describing the effect of gravity is the ratio $\kappa$ ($=g_Fm_F\mu_BB_q/mg$), which is the ratio of the magnetic potential energy to the gravitational potential energy. In our experiments $\kappa$ $<$ 13, therefore the effect of gravity in the magnetic trap is negligible. Another factor which compares the strength of rf coupling with that of the gravity is the value of ratio ($\omega/\Omega$) in comparison to $\kappa$ \cite{Morizot:2007}. If ($\omega/\Omega$) $<\kappa$, the potential becomes coupling dominant and its minimum occurs at the minimum of coupling. In our experiments $\omega/\Omega$ is $\sim$9, which ensures the trap is dominated by the coupling and not by the gravity. Without the fulfilment of conditions stated before, the gravity may result distortions in the symmetry of the trapping potentials, let it be double-well or ring. Under the influence of gravity, the  two wells in the double-well will not have the same separation when measured along different arcs on the ring. The ring trapping potential will also be modified to asymmetric ring potential favouring the accumulation of the atom cloud in the lower half of the ring in the gravitational direction. But in our experiments none of these distortions have been observed in the cloud images. Thus effect of gravity seems negligible due to appropriate parameters chosen in the experiments.

\section{Experimental Realization}\label{Experimental Realisation}

\begin{figure}
\includegraphics[width=8.6cm]{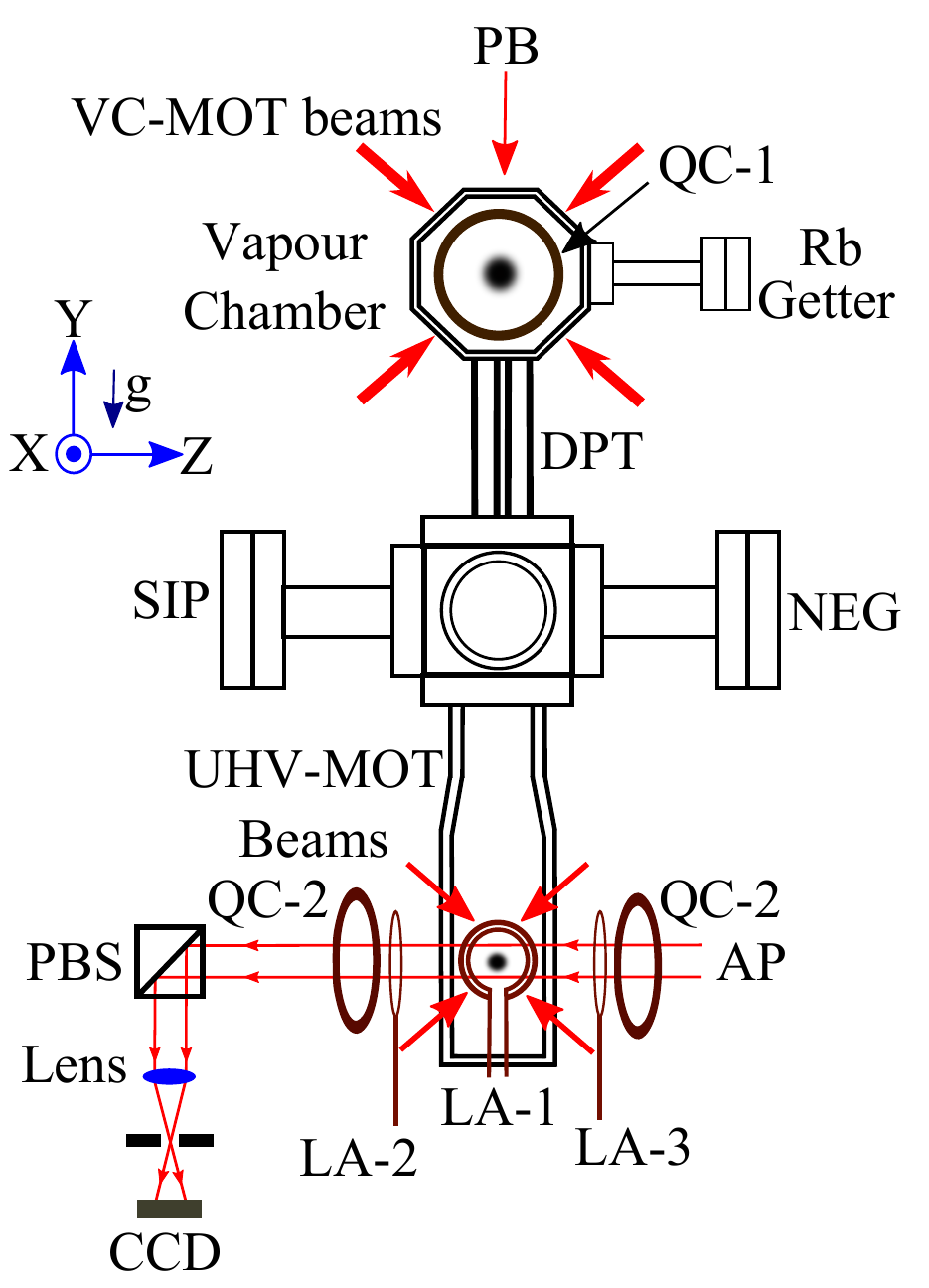}
\caption{\label{fig:setup}(Color online) Schematic diagram of the experimental setup, QC-1: Quadrupole coil for VC-MOT, QC-2: Quadrupole coil for UHV-MOT, DPT: differential pumping tube, PB: push beam, LA-1: multi-turn loop antenna for rf-dressing, LA-2: the loop antenna used for rf radiation,LA-3: pick-up loop antenna, AP: absorption probe beam. The absorption imaging optics is also shown in the diagram. }
\end{figure}

The experiments have been performed on a double magneto-optical trap (double-MOT) setup \cite{Mishra:2008,Ram:2013,Ram:2014} in which a vapour chamber MOT (VC-MOT) of $^{87}Rb$ atoms is formed in a chamber at pressure of $\sim1\times 10^{-8}$ mbar and an ultra-high vacuum MOT (UHV-MOT) is formed in a glass cell at pressure of $\sim 5\times10^{-11}$ mbar . The schematic of the setup is shown in Fig. \ref{fig:setup}. The atom cloud in the VC-MOT works as a source of atoms to load the UHV-MOT. The UHV-MOT is loaded by using a push beam focused on the VC-MOT cloud. The three cooling laser beams each with 15 mW of power are used for VC-MOT in a retro-reflecting configuration and six independent cooling beams each with 8 mW power are used for UHV-MOT. These cooling laser beams are derived from a laser beam ($\sim$ 750 mW power) which is the output of an amplifier (TA-Boosta, Toptica, Germany) seeded by an external cavity diode laser (DL-100, Toptica, Germany) acting as the master laser. Re-pumping laser beams of appropriate power are mixed with the cooling beams. Several acousto-optic modulators (AOMs) are used for control over the duration of laser beams as well as for shifting the frequency of laser beams. In addition to AOMs, the mechanical shutters are also used in the laser beams path to completely block the leakage of laser emission from AOMs during the rf dressing of the trapped atoms. This improves the life time of the trapped atoms.

We load the UHV-MOT with $\sim 2\times 10^8$ atoms and temperature $\sim$ 250 $\mu K$. The atoms in the UHV-MOT are trapped in the quadrupole magnetic trap after the temperature reduction in the compressed MOT and molasses stages. The quadrupole coils used for the UHV-MOT formation are also used for magnetic trapping of atoms in the quadrupole trap. Before the magnetic trapping, the atoms from the UHV-MOT and molasses are optically pumped to the state $5S_{1/2}|F=2, m_F=2\rangle$. This is done by applying a low magnetic field ($\sim$ 2 G, 2 ms duration) using a separate set of coils and a weak resonant optical pulse (power $\sim$ 200 $\mu$W and duration 500 $\mu$s). The optical pumping enhances the number of atoms in the trap nearly by factor of two as compared to the number achieved without optical pumping. A low-noise power supply (from TDK-Lambda) is used alongwith the appropriate switching circuits for the control of the current in the quadrupole trap coils. The rf system is assembled by connecting a rf-synthesizer (Agilent 33522A) to a broad-band rf-amplifier. The output of the amplifier is connected to a 2 cm diameter (10 turn) circular copper loop antenna (LA-1) via an impedance matching circuit. This multi-turn loop antenna LA-1 is used to apply the dressing rf-field to the atoms trapped in the quadrupole trap. The antenna LA-1 has its axis along the x-axis, whereas the symmetry axis of the quadrupole coils is along the z-axis and gravity is along the y-axis (Fig.\ref{fig:setup}). Another rf system and a single loop antenna LA-2 oriented along z-axis is used for evaporative cooling purpose. The antenna LA-3 (identical to LA-2) is used for the purpose of monitoring the frequency, amplitude and time duration of rf-radiation indirectly, as a pick-up coil. By applying a probe beam along the z-axis, derived from the cooling laser, the absorption images of the trapped cloud are recorded using a digital CCD camera (Pixelfly model USB) and appropriate 2f-imaging optics. The absorption images are processed using the Python codes. The colors in an absorption image from blue to red denote the optical density values in increasing order from low to high. 

A field programmable gate array (FPGA) card based controller system is utilized to control the various processes and sequences during the experiments. The controller system generates the pulses to trigger/switch various components like AOMs, mechanical shutters, power supplies, ccd cameras etc. The controller is operated through an industrial PC and LabVIEW software. The fast switching of current in magnetic trap coils is achieved using an IGBT-based high speed switching circuit.
 
As a first step in the experiments, VC-MOT is loaded from the back-ground Rb vapour in the vapour chamber and simultaneously atoms are transferred to the UHV-MOT for a duration of 20 s using a push beam (PB). Then the atoms in the UHV-MOT are kept in the compressed-MOT for 20 ms by increasing the axial gradient of UHV-MOT magnetic field from 6 to 8 G $cm^{-1}$. After this, the magnetic field is switched off and the atoms are transferred to an optical molasses for 5 ms, in which the temperature of the cloud decreases to $\sim$40 $\mu K$. After the optical molasses stage, the UHV-MOT laser beams are then turned off, and optical pumping is done for 500 $\mu s$. After optical pumping of the atom cloud, the current in the UHV-MOT coils (\textit{i.e.} quadrupole trap coils) is switched-on (0 to 15 A in 2ms) to trap atoms in the magnetic trap with radial gradient of 80 G $cm^{-1}$. Subsequently, the current in quadrupole trap coils is ramped slowly to increase this field gradient from 80 to 100 G $cm^{-1}$ in 1 s to form a quadrupole magnetic trap. The atom cloud in the final quadrupole trap has $\sim1\times 10^{7}$ atoms with an approximate temperature and trap lifetime $\sim$ 250 $\mu K$ and $\sim$ 18 s respectively. The rf field is switched-on after the atom cloud in the final quadrupole trap is thermalized to more than 50 ms. 

\section{Results and discussion}\label{Results}
\begin{figure}[b]
\includegraphics[width=8.6cm]{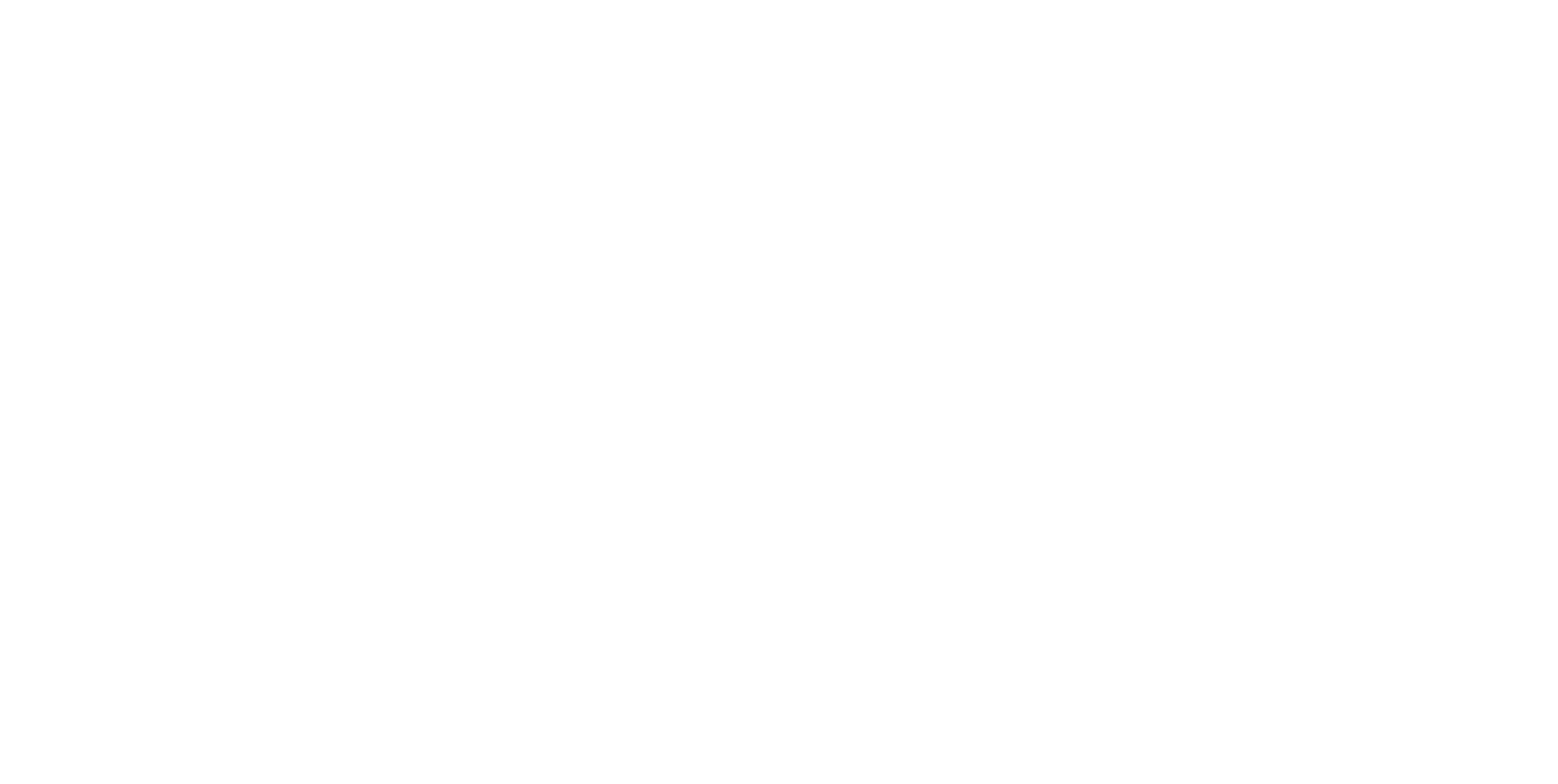}
\caption{\label{fig:image1}(Color online) Observed absorption images of the cloud (in xy-plane) after evaporative cooling in quadrupole trap: (a) before rf-dressing and (b) after rf-dressing. The curves show the profiles of optical density along the selected diameters across the image. The arrows indicate the approximate optical density at the position. Different colours in an absorption image from blue to red represent the optical density values in the order from low to high.}
\end{figure}

\begin{figure}
\includegraphics[width=8.6cm]{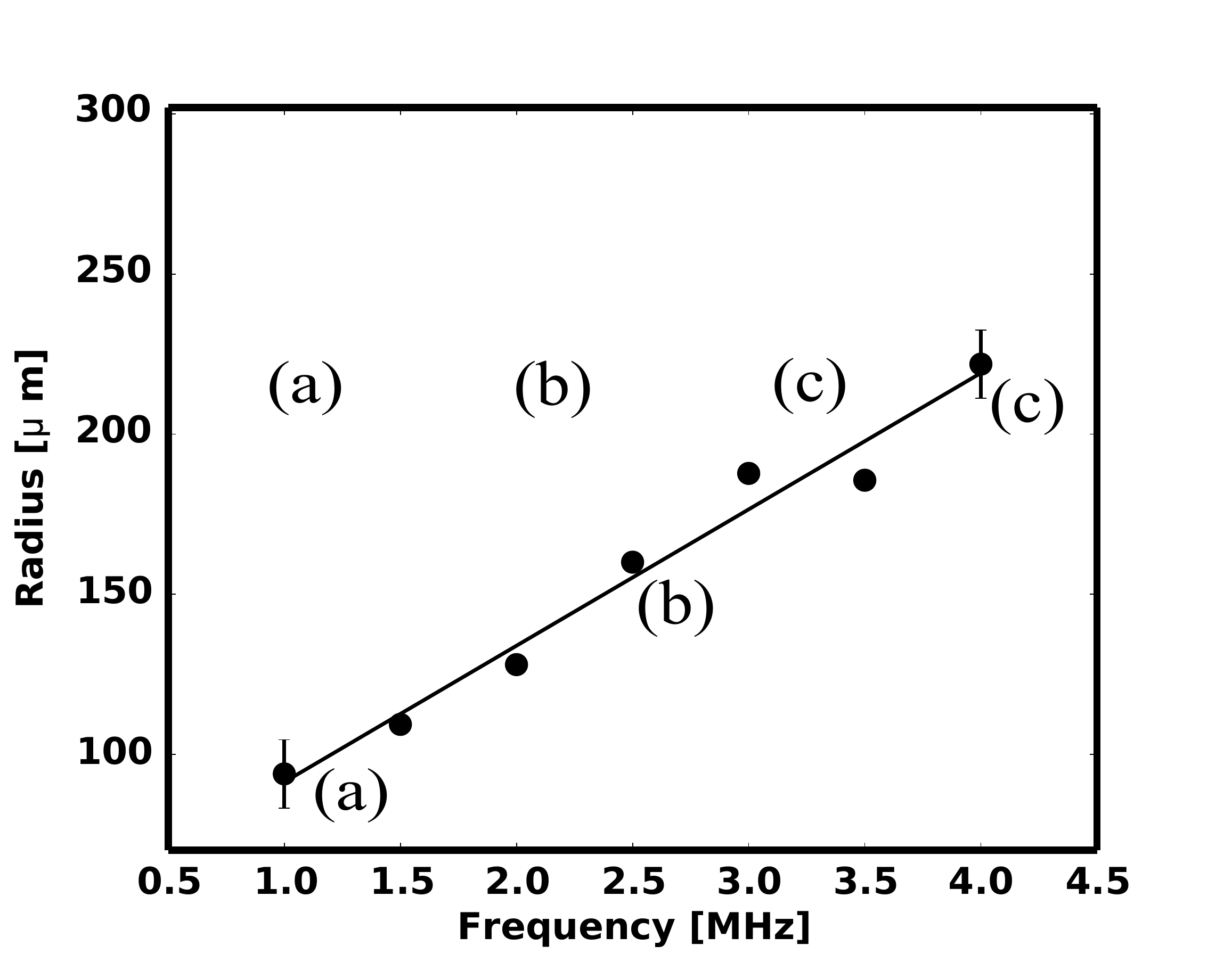}
\caption{\label{fig:radius}(Color online) Observed variation in the ring radius with frequency of the dressing rf-field. Points show the experimentally measured values and the error bars show the deviation in the value in repeated measurements. The line curve represents the theoretical prediction of radius considering experimental parameters. The inset shows the absorption images (optical density) of the atom cloud for different frequencies of dressing field.}
\end{figure}

In the experiments, the effect of the applied dressing rf-field on cold $^{87}Rb$ atoms trapped in the quadrupole magnetic trap has been studied. To start with, the atom cloud was first evaporatively cooled to reduce temperature from $\sim 250$ $\mu K$ to $\sim 20$ $\mu K$ in the quadrupole trap using a single loop copper coil antenna (LA-2) driven by a rf-synthesizer (Agilent 335220A) and its amplifier. The rf evaporation is implemented by ramping down the source frequency from 12 MHz to 3 MHz in 5 s duration. When this evaporatively cooled atom cloud in the quadrupole trap was exposed to the dressing rf-field emitted from the multi-turn antenna LA-1, the trapping of atoms in toroidal geometry was clearly observed after appropriate adjustment of power in the dressing rf-field. Figure \ref{fig:image1}(a) shows the absorption image of the cloud in the magnetic after evaporative cooling and before rf-dressing and Fig.\ref{fig:image1}(b) shows the image after rf-dressing of the atom cloud. The low density in the centre of absorption image of the cloud in the rf-dressed quadrupole trap (Fig. \ref{fig:image1}(b)) shows the trapping of atoms in the toroidal geometry. One may note here that, since dressing rf-field is linearly polarized, the rf-dressed potential in our case is expected to have a double-well structure on a circular ring (as shown in Fig. 2(a)), with two potential barriers in the azimuthal direction separating the wells. We have observed the toroidal (or ring shaped) atom cloud because the temperature ( $\sim 20$ $\mu K$) of atom cloud is higher than the height (few $\mu K$) of these azimuthal potential barriers. Due to higher temperature, atoms in the cloud can ride the barriers and move throughout the circumference of the ring. In our dressed quadrupole trap, under harmonic potential approximation near the trap minimum at ring, the ratio of axial and radial trap frequencies ($\omega_z/\omega_\rho$) is higher than 1 (estimated using approach described in \cite{Merloti:2013}). Hence the confinement along the z-direction is stronger than that in the radial direction. This indicates that the trapped cloud is quasi-2D type.

Due to non-uniform potential depth around the toroid, distribution of atom cloud is expected to be non-uniform, as observed experimentally. But the observed asymmetric cloud distribution pattern was shot to shot repeatable with negligibly small variations in the density distribution. In the measurements, the single image of the cloud was stored and analysed to retain the original density distribution intact, instead to taking multiple images and then averaging. The measurements were repeated for a number of times to check the reproducibility and derive the statistical errors. 

To estimate the radius of the toroidal ring in the absorption image of the cloud, the optical density profile along a selected diameter across the image was plotted. The measured optical density profile was fit to two Gaussian profiles  at the two ends of this diameter. The peak to peak separation between two Gaussian profiles has been used as the diameter (twice the radius) of the ring. An average over different diameters along different directions was performed to obtain the mean radius of the toroidal ring. The radius values obtained this way are shown in Fig. \ref{fig:radius} as a function of the rf frequency $\omega$. The experimental and theoretical values shown in the figure are in reasonable agreement.

\begin{figure}[b]
\includegraphics[width=8.6cm]{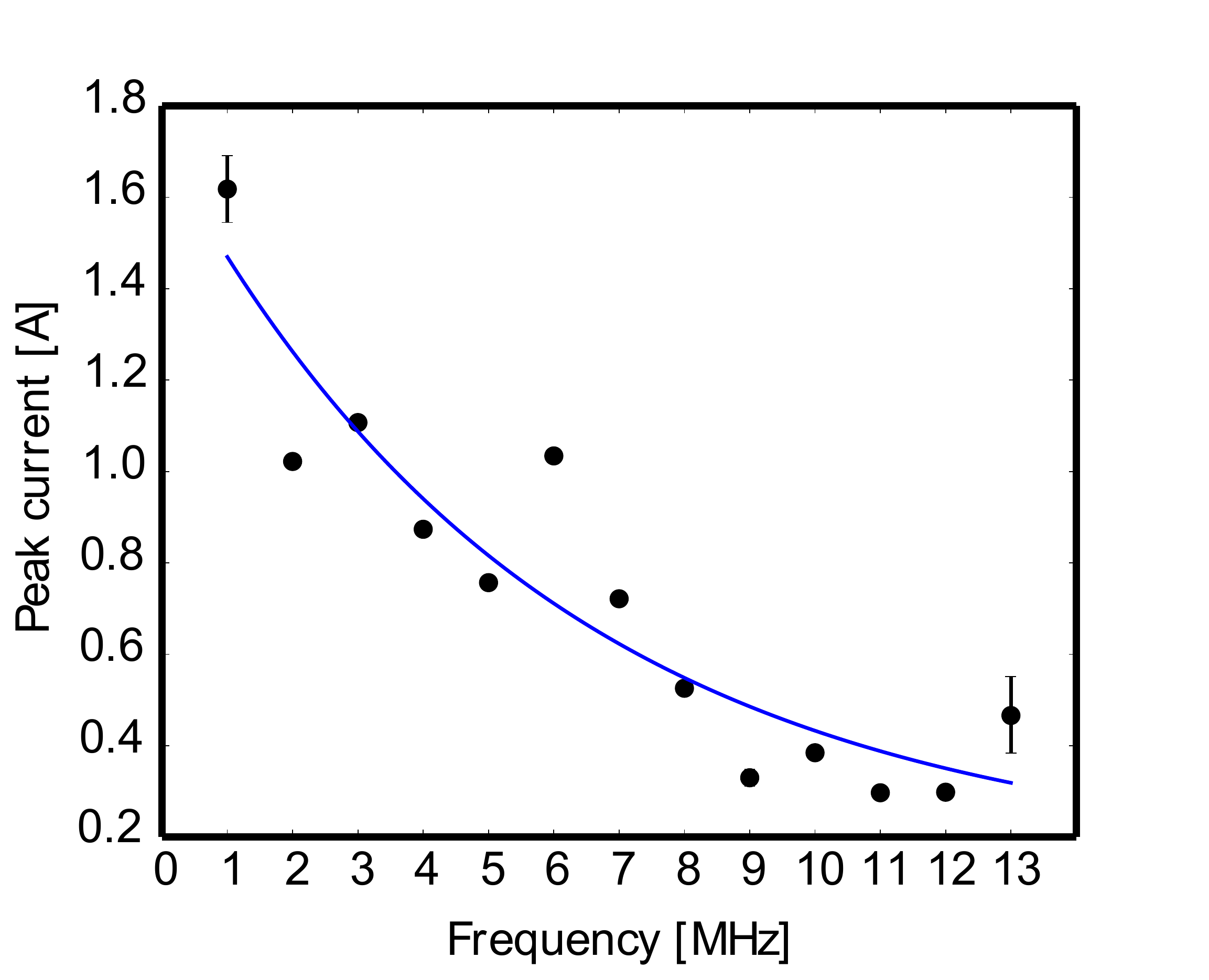}
\caption{\label{fig:response}(Color online) Peak current through the coil LA-1 as a function of frequency ($\omega/2\pi$). Vertical lines show the error bar. The continuous curve is to guide the eyes.}
\end{figure}

\begin{figure}
\includegraphics[width=8.6cm]{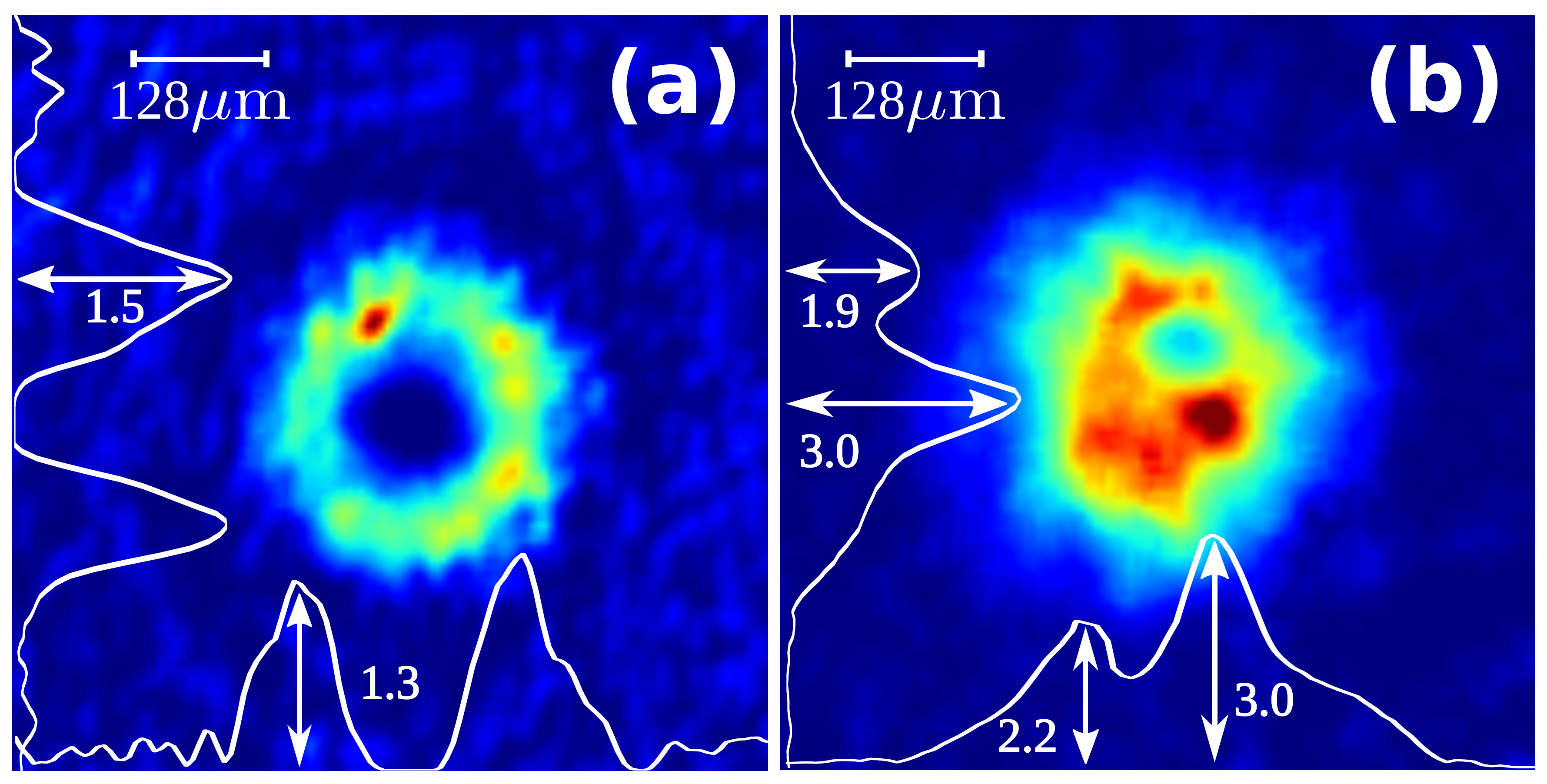}
\caption{\label{fig:image2}(Color online) (a) The absorption image of the cloud after evaporative cooling and rf-dressing using single coil LA-1 (frequency of the rf-field is ramped down from 15 MHz to 1 MHz). (b) The absorption image of the cloud when both the coils LA-1 and LA-2 are simultaneously used with frequency of the rf field ramped down from 15 MHz to 1 MHz. The curves show the variation of optical density in the images across the selected horizontal and vertical diameters. The arrows indicate the value of the measured optical density at that peak position.}
\end{figure}

In further experiments we have used single multi-turn antenna (coil LA-1) for evaporative cooling as well as for generation of the rf-dressed potential for cold atoms. The rf-frequency in this case is ramped down from 15 MHz to 1 MHz linearly in 5 s duration. The impedance matching circuit is kept resonant near the 1 MHz end of the ramp, which results in high rf-field amplitude only at lower frequencies ($\sim$1 MHz). This makes it possible to use the same antenna for evaporative cooling as well dressing of the atoms in the trap. The current variation through the coil LA-1 with frequency is shown in Fig. \ref{fig:response}. During the frequency ramp down from 15 MHz to 1 MHz, the rf-field of low amplitude (due to low current in the antenna) is radiated in the beginning which is suitable for rf evaporative cooling. In the trailing part of the ramp (near 1 MHz), a high amplitude rf-field is obtained due to multi-fold increase in the current in the antenna coil. This high amplitude rf-field converts the quadrupole trap into a rf-dressed quadrupole trap. Figure \ref{fig:image2} (a) shows a typical image of the cloud trapped in the toroidal rf-dressed potential using this single coil rf-dressing method.

Subsequently experiments have also been performed using both the coils LA-1 and LA-2 simultaneously with rf-field frequency ramped down from 15 MHz to 1 MHz. In this case the rf-field from LA-1 is stronger than that from LA-2. The presence of the z-component of rf-field, due to coil LA-2, destroys the radial symmetry of the potential and shifts the atom cloud in one arc of the toroid as shown in Fig. \ref{fig:image2}(b). This result is consistent with the results of simulations performed by incorporating z-component in the rf-field along with the linearly polarized rf-field (Fig. \ref{fig:simulation} (c)). The asymmetry arises due to inhomogeneity of the coupling strength around the toroid. As predicted by the earlier simulations \cite{Chakraborty:2014}, the position of the potential minimum can be changed by changing the value of phase $\beta$. Thus, insertion of a phase modulation unit between the two amplifiers driving LA-1 and LA-2 can be used as a technique to modulate $\beta$ and rotate the atom cloud along the toroid \cite{Chakraborty:2014,Gildemeister:2012}. Such rotation of cloud will be attempted by us using our setup. This method of rotation appears simpler than the use of a laser dipole potential to stir the cloud, which is commonly used for super-fluidity experiments.

\begin{figure}
\includegraphics[width=8.6cm]{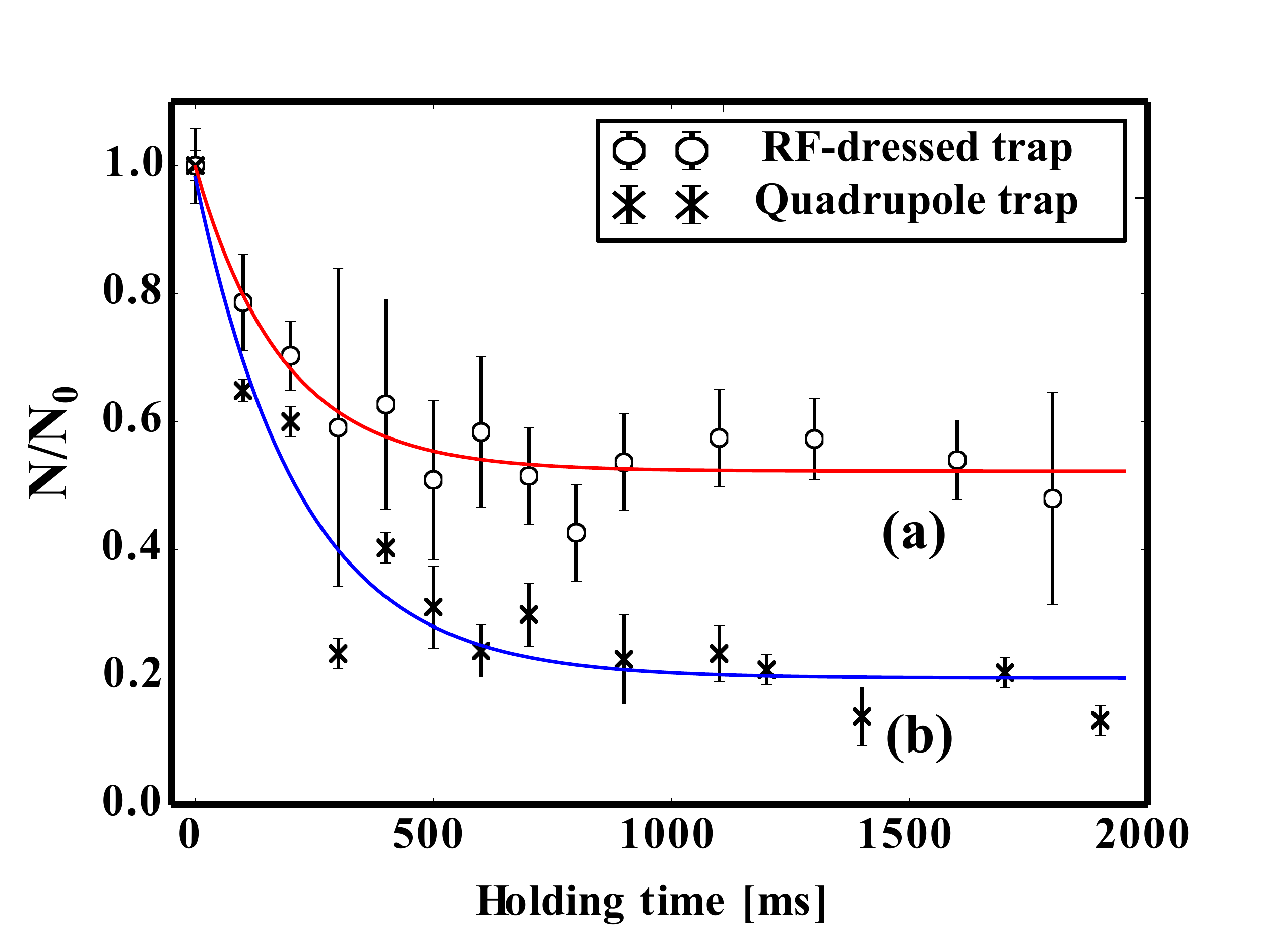}
\caption{\label{fig:holding}(Color online) Normalized number of atoms as a function of the trap holding time for (a) rf-dressed quadrupole trap and (b) bare quadrupole trap. After 1 s of holding time, the bare quadrupole trap suffers almost 30\% more loss of atoms than the rf-dressed quadrupole trap.}
\end{figure}

Next we measured the evolution in number of atoms trapped in the rf-dressed potential with time. For these measurements, the rf-dressing field (coil LA-1) was kept on at fixed frequency 1 MHz after the frequency ramp from 15 MHz to 1 MHz used for evaporative cooling and rf-dressing. With the dressing field ON, the number of atoms was measured as a function of holding time in rf-dressed trap. To compare it with bare quadrupole trap, the dressing field was switched-off after completion of the frequency ramp, and number of atoms was measured with holding time. Figure \ref{fig:holding} shows this variation in number of atoms in the rf-dressed quadrupole trap (graph (a)) and in the bare quadrupole trap (graph (b)). As shown in the Fig. \ref{fig:holding} (a), the number ($N$) drops to nearly $50\%$ of the initial number $N_0$ in 1 s. But if the dressing field is switched-off and the cloud is released to the bare magnetic trap potential, then it drops to $80\%$  (Fig.\ref{fig:holding} (b)) in the same time duration. The additional $30\%$ loss in atom number can be attributed for various loss mechanisms of the bare trap, in which Majorana transition could be the dominant one. Therefore rf dressing of the quadrupole trap reduces the losses due to Majorana transitions and hence can be useful for rf evaporative cooling to achieve degeneracy in the cold gas samples \cite{Alzar:2006,Easwaran:2010}.

\section{Conclusion}\label{Conclusion}
The trapping of the cold $^{87}Rb$ atoms in a toroidal geometry has been demonstrated in a rf-dressed quadrupole trap. It is also shown that toroidal trap can be formed using a single coil for both the purposes, evaporative cooling and rf-dressing. As predicted, by using two mutually perpendicular dressing rf-fields, the asymmetric ring trapping is also observed. The rf-dressed trap has shown lower loss rate for atoms than the bare magnetic trap. This kind of traps with further lower temperature can be used to study tunnelling and super-fluidity in low dimensions.

\section*{Acknowledgement}
We thank V. B. Tiwari for his suggestions on the manuscript and A. Srivastava for her help during the experiments. We also thank M. Lad and P. S. Bagduwal for providing the rf amplifier and A. K. Pathak, S. Tiwari and L. Jain for modifications in the control system for the experiments. A. Chakraborty acknowledges the financial support by the Homi Bhabha National Institute, India.


\begin{thebibliography}{29}
\expandafter\ifx\csname natexlab\endcsname\relax\def\natexlab#1{#1}\fi
\expandafter\ifx\csname bibnamefont\endcsname\relax
  \def\bibnamefont#1{#1}\fi
\expandafter\ifx\csname bibfnamefont\endcsname\relax
  \def\bibfnamefont#1{#1}\fi
\expandafter\ifx\csname citenamefont\endcsname\relax
  \def\citenamefont#1{#1}\fi
\expandafter\ifx\csname url\endcsname\relax
  \def\url#1{\texttt{#1}}\fi
\expandafter\ifx\csname urlprefix\endcsname\relax\def\urlprefix{URL }\fi
\providecommand{\bibinfo}[2]{#2}
\providecommand{\eprint}[2][]{\url{#2}}

\bibitem[{\citenamefont{Windpassinger and
  Sengstock}(2013)}]{Windpassinger:2013}
\bibinfo{author}{\bibfnamefont{P.}~\bibnamefont{Windpassinger}}
  \bibnamefont{and}
  \bibinfo{author}{\bibfnamefont{K.}~\bibnamefont{Sengstock}},
  \bibinfo{journal}{Reports on Progress in Physics}
  \textbf{\bibinfo{volume}{76}}, \bibinfo{pages}{086401}
  (\bibinfo{year}{2013}),
  \urlprefix\url{http://stacks.iop.org/0034-4885/76/i=8/a=086401}.

\bibitem[{\citenamefont{Hofferberth et~al.}(2007)\citenamefont{Hofferberth,
  Fischer, Schumm, Schmiedmayer, and Lesanovsky}}]{Hofferberth:2007}
\bibinfo{author}{\bibfnamefont{S.}~\bibnamefont{Hofferberth}},
  \bibinfo{author}{\bibfnamefont{B.}~\bibnamefont{Fischer}},
  \bibinfo{author}{\bibfnamefont{T.}~\bibnamefont{Schumm}},
  \bibinfo{author}{\bibfnamefont{J.}~\bibnamefont{Schmiedmayer}},
  \bibnamefont{and}
  \bibinfo{author}{\bibfnamefont{I.}~\bibnamefont{Lesanovsky}},
  \bibinfo{journal}{Phys. Rev. A} \textbf{\bibinfo{volume}{76}},
  \bibinfo{pages}{013401} (\bibinfo{year}{2007}),
  \urlprefix\url{http://link.aps.org/doi/10.1103/PhysRevA.76.013401}.

\bibitem[{\citenamefont{Merloti
  et~al.}(2013{\natexlab{a}})\citenamefont{Merloti, Dubessy, Longchambon,
  Perrin, Pottie, Lorent, and Perrin}}]{Merloti:2013}
\bibinfo{author}{\bibfnamefont{K.}~\bibnamefont{Merloti}},
  \bibinfo{author}{\bibfnamefont{R.}~\bibnamefont{Dubessy}},
  \bibinfo{author}{\bibfnamefont{L.}~\bibnamefont{Longchambon}},
  \bibinfo{author}{\bibfnamefont{A.}~\bibnamefont{Perrin}},
  \bibinfo{author}{\bibfnamefont{P.-E.} \bibnamefont{Pottie}},
  \bibinfo{author}{\bibfnamefont{V.}~\bibnamefont{Lorent}}, \bibnamefont{and}
  \bibinfo{author}{\bibfnamefont{H.}~\bibnamefont{Perrin}},
  \bibinfo{journal}{New Journal of Physics} \textbf{\bibinfo{volume}{15}},
  \bibinfo{pages}{033007} (\bibinfo{year}{2013}{\natexlab{a}}),
  \urlprefix\url{http://stacks.iop.org/1367-2630/15/i=3/a=033007}.

\bibitem[{\citenamefont{{D.S. Petrov} et~al.}(2004)\citenamefont{{D.S. Petrov},
  {D.M. Gangardt}, and {G.V. Shlyapnikov}}}]{Petrov:2004}
\bibinfo{author}{\bibnamefont{{D.S. Petrov}}},
  \bibinfo{author}{\bibnamefont{{D.M. Gangardt}}}, \bibnamefont{and}
  \bibinfo{author}{\bibnamefont{{G.V. Shlyapnikov}}}, \bibinfo{journal}{J.
  Phys. IV France} \textbf{\bibinfo{volume}{116}}, \bibinfo{pages}{5}
  (\bibinfo{year}{2004}),
  \urlprefix\url{http://dx.doi.org/10.1051/jp4:2004116001}.

\bibitem[{\citenamefont{Bloch}(1973)}]{Bloch:1973}
\bibinfo{author}{\bibfnamefont{F.}~\bibnamefont{Bloch}},
  \bibinfo{journal}{Phys. Rev. A} \textbf{\bibinfo{volume}{7}},
  \bibinfo{pages}{2187} (\bibinfo{year}{1973}),
  \urlprefix\url{http://journals.aps.org/pra/pdf/10.1103/PhysRevA.7.2187}.

\bibitem[{\citenamefont{Albiez et~al.}(2005)\citenamefont{Albiez, Gati,
  F\"olling, Hunsmann, Cristiani, and Oberthaler}}]{Albiez:2005}
\bibinfo{author}{\bibfnamefont{M.}~\bibnamefont{Albiez}},
  \bibinfo{author}{\bibfnamefont{R.}~\bibnamefont{Gati}},
  \bibinfo{author}{\bibfnamefont{J.}~\bibnamefont{F\"olling}},
  \bibinfo{author}{\bibfnamefont{S.}~\bibnamefont{Hunsmann}},
  \bibinfo{author}{\bibfnamefont{M.}~\bibnamefont{Cristiani}},
  \bibnamefont{and} \bibinfo{author}{\bibfnamefont{M.~K.}
  \bibnamefont{Oberthaler}}, \bibinfo{journal}{Phys. Rev. Lett.}
  \textbf{\bibinfo{volume}{95}}, \bibinfo{pages}{010402}
  (\bibinfo{year}{2005}),
  \urlprefix\url{http://link.aps.org/doi/10.1103/PhysRevLett.95.010402}.

\bibitem[{\citenamefont{Zobay and Garraway}(2001)}]{Zobay:2001}
\bibinfo{author}{\bibfnamefont{O.}~\bibnamefont{Zobay}} \bibnamefont{and}
  \bibinfo{author}{\bibfnamefont{B.~M.} \bibnamefont{Garraway}},
  \bibinfo{journal}{Phys. Rev. Lett.} \textbf{\bibinfo{volume}{86}},
  \bibinfo{pages}{1195} (\bibinfo{year}{2001}),
  \urlprefix\url{http://link.aps.org/doi/10.1103/PhysRevLett.86.1195}.

\bibitem[{\citenamefont{Fernholz et~al.}(2007)\citenamefont{Fernholz,
  Gerritsma, Kr\"uger, and Spreeuw}}]{Fernholz:2007}
\bibinfo{author}{\bibfnamefont{T.}~\bibnamefont{Fernholz}},
  \bibinfo{author}{\bibfnamefont{R.}~\bibnamefont{Gerritsma}},
  \bibinfo{author}{\bibfnamefont{P.}~\bibnamefont{Kr\"uger}}, \bibnamefont{and}
  \bibinfo{author}{\bibfnamefont{R.~J.~C.} \bibnamefont{Spreeuw}},
  \bibinfo{journal}{Phys. Rev. A} \textbf{\bibinfo{volume}{75}},
  \bibinfo{pages}{063406} (\bibinfo{year}{2007}),
  \urlprefix\url{http://link.aps.org/doi/10.1103/PhysRevA.75.063406}.

\bibitem[{\citenamefont{Morizot et~al.}(2006)\citenamefont{Morizot, Colombe,
  Lorent, Perrin, and Garraway}}]{Morizot:2006}
\bibinfo{author}{\bibfnamefont{O.}~\bibnamefont{Morizot}},
  \bibinfo{author}{\bibfnamefont{Y.}~\bibnamefont{Colombe}},
  \bibinfo{author}{\bibfnamefont{V.}~\bibnamefont{Lorent}},
  \bibinfo{author}{\bibfnamefont{H.}~\bibnamefont{Perrin}}, \bibnamefont{and}
  \bibinfo{author}{\bibfnamefont{B.~M.} \bibnamefont{Garraway}},
  \bibinfo{journal}{Phys. Rev. A} \textbf{\bibinfo{volume}{74}},
  \bibinfo{pages}{023617} (\bibinfo{year}{2006}),
  \urlprefix\url{http://link.aps.org/doi/10.1103/PhysRevA.74.023617}.

\bibitem[{\citenamefont{Heathcote et~al.}(2008)\citenamefont{Heathcote, Nugent,
  Sheard, and Foot}}]{Heathcote:2008}
\bibinfo{author}{\bibfnamefont{W.~H.} \bibnamefont{Heathcote}},
  \bibinfo{author}{\bibfnamefont{E.}~\bibnamefont{Nugent}},
  \bibinfo{author}{\bibfnamefont{B.~T.} \bibnamefont{Sheard}},
  \bibnamefont{and} \bibinfo{author}{\bibfnamefont{C.~J.} \bibnamefont{Foot}},
  \bibinfo{journal}{New Journal of Physics} \textbf{\bibinfo{volume}{10}},
  \bibinfo{pages}{043012} (\bibinfo{year}{2008}),
  \urlprefix\url{http://stacks.iop.org/1367-2630/10/i=4/a=043012}.

\bibitem[{\citenamefont{Amico et~al.}(2009)\citenamefont{Amico, Aghamalyan,
  Auksztol, Crepaz, Dumke, and Kwek}}]{Amico:2009}
\bibinfo{author}{\bibfnamefont{L.}~\bibnamefont{Amico}},
  \bibinfo{author}{\bibfnamefont{D.}~\bibnamefont{Aghamalyan}},
  \bibinfo{author}{\bibfnamefont{F.}~\bibnamefont{Auksztol}},
  \bibinfo{author}{\bibfnamefont{H.}~\bibnamefont{Crepaz}},
  \bibinfo{author}{\bibfnamefont{R.}~\bibnamefont{Dumke}}, \bibnamefont{and}
  \bibinfo{author}{\bibfnamefont{L.~C.} \bibnamefont{Kwek}},
  \bibinfo{journal}{Scientific Reports} \textbf{\bibinfo{volume}{4}},
  \bibinfo{pages}{4298} (\bibinfo{year}{2009}),
  \urlprefix\url{http://www.nature.com/srep/2014/140306/srep04298/full/srep04298.html}.

\bibitem[{\citenamefont{Folman et~al.}(2002)\citenamefont{Folman, Krueger,
  Schmiedmayer, Denschlag, and Henkel}}]{Folman:2002}
\bibinfo{author}{\bibfnamefont{R.}~\bibnamefont{Folman}},
  \bibinfo{author}{\bibfnamefont{P.}~\bibnamefont{Krueger}},
  \bibinfo{author}{\bibfnamefont{J.}~\bibnamefont{Schmiedmayer}},
  \bibinfo{author}{\bibfnamefont{J.}~\bibnamefont{Denschlag}},
  \bibnamefont{and} \bibinfo{author}{\bibfnamefont{C.}~\bibnamefont{Henkel}},
  \bibinfo{journal}{Adv. At. Mol. Opt. Phys.} \textbf{\bibinfo{volume}{48}},
  \bibinfo{pages}{263} (\bibinfo{year}{2002}).

\bibitem[{\citenamefont{Colombe et~al.}(2004)\citenamefont{Colombe, Knyazchyan,
  Morizot, Mercier, Lorent, and Perrin}}]{Colombe:2004}
\bibinfo{author}{\bibfnamefont{Y.}~\bibnamefont{Colombe}},
  \bibinfo{author}{\bibfnamefont{E.}~\bibnamefont{Knyazchyan}},
  \bibinfo{author}{\bibfnamefont{O.}~\bibnamefont{Morizot}},
  \bibinfo{author}{\bibfnamefont{B.}~\bibnamefont{Mercier}},
  \bibinfo{author}{\bibfnamefont{V.}~\bibnamefont{Lorent}}, \bibnamefont{and}
  \bibinfo{author}{\bibfnamefont{H.}~\bibnamefont{Perrin}},
  \bibinfo{journal}{EPL (Europhysics Letters)} \textbf{\bibinfo{volume}{67}},
  \bibinfo{pages}{593} (\bibinfo{year}{2004}),
  \urlprefix\url{http://stacks.iop.org/0295-5075/67/i=4/a=593}.

\bibitem[{\citenamefont{Lesanovsky et~al.}(2006)\citenamefont{Lesanovsky,
  Schumm, Hofferberth, Andersson, Kr\"uger, and
  Schmiedmayer}}]{Lesanovsky:2006:73}
\bibinfo{author}{\bibfnamefont{I.}~\bibnamefont{Lesanovsky}},
  \bibinfo{author}{\bibfnamefont{T.}~\bibnamefont{Schumm}},
  \bibinfo{author}{\bibfnamefont{S.}~\bibnamefont{Hofferberth}},
  \bibinfo{author}{\bibfnamefont{L.~M.} \bibnamefont{Andersson}},
  \bibinfo{author}{\bibfnamefont{P.}~\bibnamefont{Kr\"uger}}, \bibnamefont{and}
  \bibinfo{author}{\bibfnamefont{J.}~\bibnamefont{Schmiedmayer}},
  \bibinfo{journal}{Phys. Rev. A} \textbf{\bibinfo{volume}{73}},
  \bibinfo{pages}{033619} (\bibinfo{year}{2006}),
  \urlprefix\url{http://link.aps.org/doi/10.1103/PhysRevA.73.033619}.

\bibitem[{\citenamefont{Morizot et~al.}(2007)\citenamefont{Morizot, Alzar,
  Pottie, Lorent, and Perrin}}]{Morizot:2007}
\bibinfo{author}{\bibfnamefont{O.}~\bibnamefont{Morizot}},
  \bibinfo{author}{\bibfnamefont{C.~L.~G.} \bibnamefont{Alzar}},
  \bibinfo{author}{\bibfnamefont{P.-E.} \bibnamefont{Pottie}},
  \bibinfo{author}{\bibfnamefont{V.}~\bibnamefont{Lorent}}, \bibnamefont{and}
  \bibinfo{author}{\bibfnamefont{H.}~\bibnamefont{Perrin}},
  \bibinfo{journal}{Journal of Physics B: Atomic, Molecular and Optical
  Physics} \textbf{\bibinfo{volume}{40}}, \bibinfo{pages}{4013}
  (\bibinfo{year}{2007}),
  \urlprefix\url{http://stacks.iop.org/0953-4075/40/i=20/a=004}.

\bibitem[{\citenamefont{Sherlock et~al.}(2011)\citenamefont{Sherlock,
  Gildemeister, Owen, Nugent, and Foot}}]{Sherlock:2011}
\bibinfo{author}{\bibfnamefont{B.~E.} \bibnamefont{Sherlock}},
  \bibinfo{author}{\bibfnamefont{M.}~\bibnamefont{Gildemeister}},
  \bibinfo{author}{\bibfnamefont{E.}~\bibnamefont{Owen}},
  \bibinfo{author}{\bibfnamefont{E.}~\bibnamefont{Nugent}}, \bibnamefont{and}
  \bibinfo{author}{\bibfnamefont{C.~J.} \bibnamefont{Foot}},
  \bibinfo{journal}{Phys. Rev. A} \textbf{\bibinfo{volume}{83}},
  \bibinfo{pages}{043408} (\bibinfo{year}{2011}),
  \urlprefix\url{http://link.aps.org/doi/10.1103/PhysRevA.83.043408}.

\bibitem[{\citenamefont{Lesanovsky and von Klitzing}(2007)}]{Lesanovsky:2007}
\bibinfo{author}{\bibfnamefont{I.}~\bibnamefont{Lesanovsky}} \bibnamefont{and}
  \bibinfo{author}{\bibfnamefont{W.}~\bibnamefont{von Klitzing}},
  \bibinfo{journal}{Phys. Rev. Lett.} \textbf{\bibinfo{volume}{99}},
  \bibinfo{pages}{083001} (\bibinfo{year}{2007}),
  \urlprefix\url{http://link.aps.org/doi/10.1103/PhysRevLett.99.083001}.

\bibitem[{\citenamefont{Merloti
  et~al.}(2013{\natexlab{b}})\citenamefont{Merloti, Dubessy, Longchambon,
  Olshanii, and Perrin}}]{Merloti:R:2013}
\bibinfo{author}{\bibfnamefont{K.}~\bibnamefont{Merloti}},
  \bibinfo{author}{\bibfnamefont{R.}~\bibnamefont{Dubessy}},
  \bibinfo{author}{\bibfnamefont{L.}~\bibnamefont{Longchambon}},
  \bibinfo{author}{\bibfnamefont{M.}~\bibnamefont{Olshanii}}, \bibnamefont{and}
  \bibinfo{author}{\bibfnamefont{H.}~\bibnamefont{Perrin}},
  \bibinfo{journal}{Phys. Rev. A.} \textbf{\bibinfo{volume}{88}},
  \bibinfo{pages}{061603} (\bibinfo{year}{2013}{\natexlab{b}}).

\bibitem[{\citenamefont{Morizot et~al.}(2008)\citenamefont{Morizot,
  Longchambon, Kollengode~Easwaran, Dubessy, Knyazchyan, Pottie, Lorent, and
  Perrin}}]{Morizot:2008}
\bibinfo{author}{\bibfnamefont{O.}~\bibnamefont{Morizot}},
  \bibinfo{author}{\bibfnamefont{L.}~\bibnamefont{Longchambon}},
  \bibinfo{author}{\bibfnamefont{R.}~\bibnamefont{Kollengode~Easwaran}},
  \bibinfo{author}{\bibfnamefont{R.}~\bibnamefont{Dubessy}},
  \bibinfo{author}{\bibfnamefont{E.}~\bibnamefont{Knyazchyan}},
  \bibinfo{author}{\bibfnamefont{P.-E.} \bibnamefont{Pottie}},
  \bibinfo{author}{\bibfnamefont{V.}~\bibnamefont{Lorent}}, \bibnamefont{and}
  \bibinfo{author}{\bibfnamefont{H.}~\bibnamefont{Perrin}},
  \bibinfo{journal}{The European Physical Journal D}
  \textbf{\bibinfo{volume}{47}}, \bibinfo{pages}{209} (\bibinfo{year}{2008}),
  ISSN \bibinfo{issn}{1434-6060},
  \urlprefix\url{http://dx.doi.org/10.1140/epjd/e2008-00050-2}.

\bibitem[{\citenamefont{Gildemeister et~al.}(2010)\citenamefont{Gildemeister,
  Nugent, Sherlock, Kubasik, Sheard, and Foot}}]{Gildemeister:2010}
\bibinfo{author}{\bibfnamefont{M.}~\bibnamefont{Gildemeister}},
  \bibinfo{author}{\bibfnamefont{E.}~\bibnamefont{Nugent}},
  \bibinfo{author}{\bibfnamefont{B.~E.} \bibnamefont{Sherlock}},
  \bibinfo{author}{\bibfnamefont{M.}~\bibnamefont{Kubasik}},
  \bibinfo{author}{\bibfnamefont{B.~T.} \bibnamefont{Sheard}},
  \bibnamefont{and} \bibinfo{author}{\bibfnamefont{C.~J.} \bibnamefont{Foot}},
  \bibinfo{journal}{Phys. Rev. A} \textbf{\bibinfo{volume}{81}},
  \bibinfo{pages}{031402} (\bibinfo{year}{2010}),
  \urlprefix\url{http://link.aps.org/doi/10.1103/PhysRevA.81.031402}.

\bibitem[{\citenamefont{Chakraborty and Mishra}(2014)}]{Chakraborty:2014}
\bibinfo{author}{\bibfnamefont{A.}~\bibnamefont{Chakraborty}} \bibnamefont{and}
  \bibinfo{author}{\bibfnamefont{S.~R.} \bibnamefont{Mishra}},
  \bibinfo{journal}{J. of the Kor. Phys. Soc.} \textbf{\bibinfo{volume}{65}},
  \bibinfo{pages}{1324} (\bibinfo{year}{2014}),
  \urlprefix\url{http://link.springer.com/article/10.3938%2Fjkps.65.1324}.

\bibitem[{\citenamefont{Schumm et~al.}(2005)\citenamefont{Schumm, Hofferberth,
  Anderson, Widermuth, Groth, Bar-Joseph, Schmiedmayer, and
  Kruger}}]{Schumm:2005}
\bibinfo{author}{\bibfnamefont{T.}~\bibnamefont{Schumm}},
  \bibinfo{author}{\bibfnamefont{S.}~\bibnamefont{Hofferberth}},
  \bibinfo{author}{\bibfnamefont{L.~M.} \bibnamefont{Anderson}},
  \bibinfo{author}{\bibfnamefont{S.}~\bibnamefont{Widermuth}},
  \bibinfo{author}{\bibfnamefont{S.}~\bibnamefont{Groth}},
  \bibinfo{author}{\bibfnamefont{I.}~\bibnamefont{Bar-Joseph}},
  \bibinfo{author}{\bibfnamefont{J.}~\bibnamefont{Schmiedmayer}},
  \bibnamefont{and} \bibinfo{author}{\bibfnamefont{P.}~\bibnamefont{Kruger}},
  \bibinfo{journal}{Nat. Phys.} \textbf{\bibinfo{volume}{1}},
  \bibinfo{pages}{57} (\bibinfo{year}{2005}),
  \urlprefix\url{http://www.nature.com/nphys/journal/v1/n1/full/nphys125.html}.

\bibitem[{\citenamefont{Heathcote}(2007)}]{Heathcote:thesis}
\bibinfo{author}{\bibfnamefont{W.}~\bibnamefont{Heathcote}}, Ph.D. thesis,
  \bibinfo{school}{University of Oxford} (\bibinfo{year}{2007}).

\bibitem[{\citenamefont{Mishra et~al.}(2008)\citenamefont{Mishra, Ram, Tiwari,
  and Mehendale}}]{Mishra:2008}
\bibinfo{author}{\bibfnamefont{S.~R.} \bibnamefont{Mishra}},
  \bibinfo{author}{\bibfnamefont{S.~P.} \bibnamefont{Ram}},
  \bibinfo{author}{\bibfnamefont{S.~K.} \bibnamefont{Tiwari}},
  \bibnamefont{and} \bibinfo{author}{\bibfnamefont{S.~C.}
  \bibnamefont{Mehendale}}, \bibinfo{journal}{Phys. Rev. A}
  \textbf{\bibinfo{volume}{77}}, \bibinfo{pages}{065402}
  (\bibinfo{year}{2008}),
  \urlprefix\url{http://link.aps.org/doi/10.1103/PhysRevA.77.065402}.

\bibitem[{\citenamefont{Ram et~al.}(2013)\citenamefont{Ram, Tiwari, Mishra, and
  Rawat}}]{Ram:2013}
\bibinfo{author}{\bibfnamefont{S.~P.} \bibnamefont{Ram}},
  \bibinfo{author}{\bibfnamefont{S.~K.} \bibnamefont{Tiwari}},
  \bibinfo{author}{\bibfnamefont{S.~R.} \bibnamefont{Mishra}},
  \bibnamefont{and} \bibinfo{author}{\bibfnamefont{H.~S.} \bibnamefont{Rawat}},
  \bibinfo{journal}{Review of Scientific Instruments}
  \textbf{\bibinfo{volume}{84}}, \bibinfo{pages}{073102}
  (\bibinfo{year}{2013}),
  \urlprefix\url{http://scitation.aip.org/content/aip/journal/rsi/84/7/10.1063/1.4812339}.

\bibitem[{\citenamefont{Ram et~al.}(2014)\citenamefont{Ram, Mishra, Tiwari, and
  Rawat}}]{Ram:2014}
\bibinfo{author}{\bibfnamefont{S.~P.} \bibnamefont{Ram}},
  \bibinfo{author}{\bibfnamefont{S.~R.} \bibnamefont{Mishra}},
  \bibinfo{author}{\bibfnamefont{S.~K.} \bibnamefont{Tiwari}},
  \bibnamefont{and} \bibinfo{author}{\bibfnamefont{H.~S.} \bibnamefont{Rawat}},
  \bibinfo{journal}{Journal of the Korean Physical Society}
  \textbf{\bibinfo{volume}{65}}, \bibinfo{pages}{462} (\bibinfo{year}{2014}),
  ISSN \bibinfo{issn}{0374-4884},
  \urlprefix\url{http://dx.doi.org/10.3938/jkps.65.462}.

\bibitem[{\citenamefont{Gildemeister et~al.}(2012)\citenamefont{Gildemeister,
  Sherlock, and Foot}}]{Gildemeister:2012}
\bibinfo{author}{\bibfnamefont{M.}~\bibnamefont{Gildemeister}},
  \bibinfo{author}{\bibfnamefont{B.~E.} \bibnamefont{Sherlock}},
  \bibnamefont{and} \bibinfo{author}{\bibfnamefont{C.~J.} \bibnamefont{Foot}},
  \bibinfo{journal}{Phys. Rev. A} \textbf{\bibinfo{volume}{85}},
  \bibinfo{pages}{053401} (\bibinfo{year}{2012}),
  \urlprefix\url{http://link.aps.org/doi/10.1103/PhysRevA.85.053401}.

\bibitem[{\citenamefont{Garrido~Alzar et~al.}(2006)\citenamefont{Garrido~Alzar,
  Perrin, Garraway, and Lorent}}]{Alzar:2006}
\bibinfo{author}{\bibfnamefont{C.~L.} \bibnamefont{Garrido~Alzar}},
  \bibinfo{author}{\bibfnamefont{H.}~\bibnamefont{Perrin}},
  \bibinfo{author}{\bibfnamefont{B.~M.} \bibnamefont{Garraway}},
  \bibnamefont{and} \bibinfo{author}{\bibfnamefont{V.}~\bibnamefont{Lorent}},
  \bibinfo{journal}{Phys. Rev. A} \textbf{\bibinfo{volume}{74}},
  \bibinfo{pages}{053413} (\bibinfo{year}{2006}),
  \urlprefix\url{http://link.aps.org/doi/10.1103/PhysRevA.74.053413}.

\bibitem[{\citenamefont{Easwaran et~al.}(2010)\citenamefont{Easwaran,
  Longchambon, Pottie, Lorent, Perrin, and Garraway}}]{Easwaran:2010}
\bibinfo{author}{\bibfnamefont{R.~K.} \bibnamefont{Easwaran}},
  \bibinfo{author}{\bibfnamefont{L.}~\bibnamefont{Longchambon}},
  \bibinfo{author}{\bibfnamefont{P.-E.} \bibnamefont{Pottie}},
  \bibinfo{author}{\bibfnamefont{V.}~\bibnamefont{Lorent}},
  \bibinfo{author}{\bibfnamefont{H.}~\bibnamefont{Perrin}}, \bibnamefont{and}
  \bibinfo{author}{\bibfnamefont{B.~M.} \bibnamefont{Garraway}},
  \bibinfo{journal}{Journal of Physics B: Atomic, Molecular and Optical
  Physics} \textbf{\bibinfo{volume}{43}}, \bibinfo{pages}{065302}
  (\bibinfo{year}{2010}),
  \urlprefix\url{http://stacks.iop.org/0953-4075/43/i=6/a=065302}.

\end{thebibliography}

\end{document}